\documentclass{aa}
\usepackage{graphics}
\begin{document}
\thesaurus{11        % A&A Section 11. Galaxies
          (11.19.3;  % Galaxies : starburst,
           11.16.1)}  % Galaxies : photometry.
\title{Photometric studies of some starburst galaxies.}
\author{A. Chitre and U.C. Joshi}
\offprints{A. Chitre}
\institute{Physical Research Laboratory, Navrangpura,
Ahmedabad, Gujarat, India-380 009. 
\\email: chitre@prl.ernet.in}
\date{Received ; accepted }
\maketitle

\begin{abstract}
We present the results of a detailed morphological analysis of ten
starburst galaxies selected from the Markarian catalogue of uv-excess
objects. CCD surface photometry of these galaxies was carried out based
on observations made in B,
V (Johnson) and R, I (Kron-Cousins) band passes. We present the radial
variations of surface brightness, ellipticity, position angle and the
colour indices for each galaxy obtained using ellipse fitting isophotal
analysis. The residual images constructed for extracting the fine
structure are also presented. A variety of morphological types are found
to host the starburst phenomenon. The star formation activity is not
confined to the nuclear region alone, but it also occurs at various
locations in the galaxy and is seen as clumpy regions. The colour index and the
residual images are used for deriving information about the sites of
enhanced star formation activity and the triggers of the starburst.
The luminosity profiles show an exponential behaviour in the outer
region. The disk scale lengths and the half-light radii are derived.
The contribution of the burst component has been estimated and the
colours of the burst component are presented. 
Strong isophotal twisting is detected in all the S0 and E
galaxies: Mrk 1002, Mrk
1308 and Mrk 14, in the sample. This is accompanied by boxiness in some
cases, suggesting that a merger is responsible for the
starburst activity in these galaxies. In case of isolated spirals, a bar
or a central oval distortion appear to be the likely trigger for the
starburst. 
\end{abstract}

\section{Introduction} 

The term "starburst" was first coined by Weedman (\cite{weed}) to
describe galaxies experiencing episodes of star formation that are too
intense to be sustained over the lifetime of the galaxy. These bursts of
star formation produce 10$^7$-10$^9$ M$_\odot$ of OB stars (Balzano
\cite{balz}) which are the dominant source of the galaxy's luminosity.
Since OB stars emit copiously in the ultraviolet, many starburst galaxies
were detected in Markarian et al. (\cite{mark}) survey of uv-excess objects 
and references therein. These
lists form one of the largest databases of optically selected starburst
galaxies. Huchra (\cite{huchra}) conducted a detailed aperture
photometric study of a large number of these galaxies. A
spectrophotometric survey was conducted by Balzano (\cite{balz}) wherein
she studied the spectra of the nuclear regions in the starburst galaxies.
Larson \& Tinsley (\cite{lars}) were the first to put forward the idea
that "bursts of star formation" could be triggered by tidal forces in
interacting galaxies. Combes et al. (\cite{comb})
showed that in the presence of non-axisymmetric potentials like bars or
companions, gravitational torques are capable of driving the gas in a
galaxy into the central regions. This gas may pile-up at the ILR, if it
is present or proceed right up to the nuclear region where it may fragment
to form ring-shaped or nuclear starbursts. Interactions may also induce
extranuclear starbursts as is seen in the H$\alpha$ imaging study of
Garcia-Barreto et al. (\cite{garcia}).

Dynamical disturbances experienced by the galaxy are reflected in the
morphology of the galaxy in the form of peculiar structures. Tidal tails,
bridges, asymmetric outer envelopes are some manifestations of the
perturbing processes (Combes \cite{combes} and references therein). These disturbances are
possible triggers of the starburst phenomenon. Interactions, mergers and
the presence of a bar were proposed as the most likely triggers for the
starburst. However, it has been observed that a few starburst galaxies
show neither visible signs of disturbances in the form of peculiar
morphologies nor the presence of a bar in their direct images. A detailed
study of the underlying galaxy is especially important in such cases to
probe the cause of the starburst. An inspection of the Markarian
sample of starburst galaxies reveals that though the starburst phenomenon
is predominantly found in spiral galaxies, it is not confined to spirals
alone and the sample contains a large number of S0 and elliptical galaxies, in
addition to the ones that show too peculiar a morphology to be classified
properly. 
A morphological study of a sample of starburst galaxies through several
spectral bands is expected to help shed some light on the conditions
leading to the onset of a starburst and the environment of the
underlying galaxies. The general structure of the underlying galaxy can
be studied through techniques like ellipse fitting. This technique is
especially useful in studying isophotal twists, the
presence of bars, rings, shells, etc. in the galaxy.
In recent years, studies conducted by Wozniak et
al. (\cite{woz95}) and Jungwiert et al. (\cite{jung}) on samples of
spiral galaxies using this technique have been useful in detecting
underlying structures like oval
distortions, shells, rings and hitherto undetected bars. 
This technique can be used to study the starburst galaxies in a similar
manner. The fine structure extracted from the isophotal analysis can be
interpreted to give valuable information about the underlying galaxy and
the nature of the disturbances experienced by it (Kenney et al.
\cite{kenney}). In the present study, a sample of starburst galaxies was
observed in BVRI bands and we present results on ten galaxies.

\section{Observations and data reduction} 

The sample presented in this paper is a part of a sample of starburst
galaxies selected from the Markarian catalog. Galaxies with m$_v$ brighter than
$14^m.5$ and angular sizes greater than 30\arcsec\ were
selected. No consideration was made for any particular morphological type
and the sample is unbiased towards the morphology of the galaxy. In this paper,
results based on the photometric studies of ten galaxies from the sample are
presented. These ten galaxies
are listed in Table 1., along with their coordinates, morphological
types, distances (H$_0$ = 75 kms$^{-1}$Mpc$^{-1}$ has been adopted
throughout), total exposure times in each band, the derived ellipticity,
inclination and position angle.

The observations were made at the Cassegrain focus of the 1.2m Gurushikhar
telescope employing a thinned back illuminated Tektronix 1024$\times$1024 CCD
chip. The observations were carried out under photometric conditions with
typical seeing of 1.5 \arcsec\ . Binning of 2$\times$2 was employed before
recording the images to increase the signal-to-noise ratio of the
measurements and keeping in mind the data storage requirements. The final
resolution was 0$^\prime$$^\prime$.634 /pixel which is sufficient to sample
the PSF appropriately. The field of view provided by the chip was
5$^\prime$.2$\times$5$^\prime$.2 . Broad band Johnson's B and V and
Kron-Cousin's R and I filters were used.
About 3-4 exposures were taken in
each of the photometric bands. The galaxy Mrk 87 was observed during two runs
as the signal in the B band was found to be inadequate after the first observing run.
Twilight flats were taken and median
filtered to construct the master flats. The data was reduced using
IRAF
\footnote{IRAF is distributed by National Optical Astronomy
Observatories, which is operated by the Association of Universities Inc.
(AURA) under cooperative agreement with the National Science Foundation,
USA.} 
on the IBM-6000 RISC system at PRL, Ahmedabad. Standard procedures
of bias subtraction, flat-fielding and cosmic ray removal were done on the
raw images. For the photometric calibrations of the BVRI images, stars
from Landolt (\cite{landolt}) were observed. The photometry of the standard
stars was carried out using the DAOPHOT task in IRAF. Twenty
standard stars from Landolt having $8^m.0 < V< 15^m.0 $ and $ -0.20<B-V<
+1.00 $ were observed. Instrumental magnitudes were converted to standard
magnitudes using the following equations obtained after a least squares
fitting.
\begin{eqnarray}
(B-V) & =& (b-v)-0.48
\nonumber\\
(V-R) &=& 1.05(v-r)-0.18+0.05(B-V)
\nonumber\\
(R-I) &=& 0.94(r-i)+0.36+0.04(V-R)
\nonumber
\end{eqnarray}

where lower cases denote instrumental values and upper cases correspond to
the standard values. All the magnitudes and colours presented in this paper are in the B and V
Johnson's and R and I Kron Cousin's system which is similar to the one used by
Landolt to obtain the magnitudes and colours for the standard stars in his
catalogue. Corrections for airmass were applied using the coefficients
obtained from the transformation equations. The typical errors in the
transformed magnitudes are 0.03 in B, 0.03 in V, 0.02 in R and 0.04 in I.
The galaxy images were shifted and co-added to improve the signal-to-noise
ratio. The sky background was computed from the mode of the histogram of
the image. The field of view was large enough compared to the program
galaxies so that the histograms were sky dominated. The galaxy images were
sky subtracted and subsequently scaled for 1 second exposure.

\section{Photometric accuracy}
In our final images, the noise per pixel has a standard deviation that
corresponds to 
24.5, 23.9, 23.8 and
23.0 mag/\arcsec$^{2}$ in B, V,
R and I respectively. We used published
aperture photometry data compiled by Prugniel \& Heraudeau (\cite{prug}) to test the accuracy of
our values. Synthetic aperture photometry was carried out
on the galaxies for which aperture photometry data exists in the literature and the results of the
comparison are presented in Table 2. Our values agree very well with the values derived by other workers for most of the galaxies
confirming the accuracy of our photometry. Our values for Mrk 14 and Mrk 87
are brighter as compared to the values obtained by Huchra. In case of Mrk 87,
the difference is nearly the same within the limits of the errors and is
most probably a result of an error  in the determination of the zero-point for this observation. Mrk 213, Mrk 781 and Mrk 1379 were observed on the same night as Mrk 14. If the discrepancy in our values and those of Huchra would have been the result of an error in the zero-point determined from the transformations,
the same systematic error would also have appeared for the other galaxies observed on the same night. However, no such effect is seen. Secondly, our values are brighter than those quoted in the literature which rules out inadequate exposure time as a cause of the discrepancy. Therefore, the difference appears to be
genuine. One probable explanation for this could be a variability in the
output luminosity of this source. Optical variability may be expected in young
starbursts due to the appearance of supernovae (Terlevich {\cite{terl}).
In CCD photometry, the center of the galaxy can be determined with an accuracy of about
one-tenth of an arcsec, while it is not possible to do so in aperture
photometry. Small centering errors can also give rise to a discrepancy between
the magnitudes derived from CCD and from aperture photometry, especially for small apertures.

\begin{table*}\caption[]{Sample of starburst galaxies}
\begin{tabular}{llllllllllll}
\hline
GALAXY & $\alpha$ & $\delta$ & TYPE$^{\mathrm{a}}$ &Distance&
\multicolumn{4}{l}{exposure time (minutes)} & ellipticity & inc & P.A. \\
 & (2000) & (2000) &  & Mpc&B & V & R & I & & \degr & \degr \\
\hline
Mrk 14 & 08:10:59.1 & 72:47:41 & S0?&42&23&11&8&10&0.2&41&165\\
Mrk 87 & 08:21:41.2 & 73:59:23 & SB0/a&37&50&6&6&6&0.5&65.5&60\\
Mrk 213 & 12:31:20.5 & 57:57:47 & SBa&42&25&12&11&11&0.6&72.5&141\\
Mrk 363 & 01:50:58.0 & 21:59:50 & Scp&39&17&6&6&5.3&0.25&45.5&171\\
Mrk 449 & 13:11:30.8 & 36:16:52 & Sap&15&15&7.5&7.5&10&0.7&80&82\\
Mrk 743 & 11:38:12.9 & 12:06:43 & E0p&13&13.3&7.5&7&6&-&-&-\\
Mrk 781 & 12:53:50.7 & 09:42:33 & SBc&37&13.3&10&8.6&6&0.1&30&35\\
Mrk 1002 & 01:37:17.5 & 05:52:38 & E1&42&20&9.16&5&5&0.26&46&163\\
Mrk 1308 & 11:54:12.2 & 00:08:11 & S0&14&13.3&7.5&6&8&0.1&30&20\\
Mrk 1379 & 14:17:40.3 & -07:25:03 & SBbc dbl&37&16.6&6&6&8.6&0.45&61.5&51\\
\hline\end{tabular}
\begin{list}{}{}
\item[$^{\mathrm{a}}$] Markarian catalogue (downloaded from CDS)
\end{list}
\end{table*}
\begin{table}\caption[]{Comparison with aperture photometry from the literature}
\begin{tabular}{lclllll}
\hline
Galaxy&aperture&\multicolumn{2}{l}{our values}&\multicolumn{2}{l}{literature}&re
ference\\
Mrk&\arcsec &V&B-V&V&B-V&\\
\hline
14&15&14.67&0.42&14.81&0.42&H\\
&29&14.27&0.50&14.53&0.42&H\\
87&15&14.42&0.80&14.53&0.86&H\\
&29&13.81&0.84&14.00&0.93&H\\
&57&13.16&0.81&13.35&0.90&H\\
213&57&12.83&0.58&12.80&0.62&H\\
363&24&14.02&0.53&14.03&0.57&H\\
&38&13.86&0.54&13.87&0.57&H\\
449&81&12.97&0.66&12.99&0.66&H\\
743&23&13.82&0.34&13.95&0.34&M\\
1308&17&14.11&0.40&14.19&0.40&HM\\
&24&13.85&0.46&13.86&0.51&HM\\
1379&15.4&13.6&0.61&13.67&0.58&GHJ\\
&26.8&13.21&0.63&13.44&0.65&ARK\\
&30&13.13&0.63&13.16&0.61&GRI\\
&37&13.01&0.64&13.02&0.63&GRI\\
&39.6&12.97&0.64&13.03&0.62&GHJ\\
&67.3&12.62&0.63&12.68&0.61&GHJ\\
\hline
\end{tabular}
\begin{list}{}{}
\item[] H : Huchra (\cite{huch}); M : Mc Clure \& van den Bergh (\cite{mcv}); HM
 : Humay \& Maza (\cite{hum}); GHJ: Griersmith et al. (\cite{ghj}); ARK : Arkhip
ova (\cite{arkhi}) ; GRI: Griersmith (\cite{gri})
\end{list}
\end{table}

\begin{figure*}
\vspace{1cm}
\caption[]{Clockwise from top left, the B and I band isophotes, the residual
image and
the (B-I) image. The contours are plotted between 19-23 mag/\arcsec$^{2}$
at intervals of 0.5 mag/\arcsec$^{2}$ for the B band and between
17-22 mag/\arcsec$^{2}$ with the same interval for the I band. The same order is followed in the figures (a) to (j). For (B-I) images, dark is blue and white is red.
For Mrk 87, the B and I images are presented instead of
the isophotes to depict the inner ring clearly. North is at the the top and
east is to the left. Residual images in the R band are given only for
those galaxies which show some fine structure.}
\label{img}
\end{figure*}
\begin{figure*}
\vspace{1cm}
\caption[]{Radial variation of surface brightness, ellipticity (e) and position
angle (P.A.) in B,V,R,I from left to right.}
\label{sep}
\end{figure*}
\section{Surface photometry}
A direct inspection of the broad band images and the contour maps shows
that many of these galaxies exhibit morphological peculiarities. To
conduct a spatially resolved study of the stellar populations and the
distribution of dust, we constructed (B-V), (V-R) and (B-I) colour maps.  
The colour maps were constructed after matching the PSFs in the two bands
to avoid artefacts. This was achieved by degrading the better of the PSFs
in the two bands by using a gaussian smoothing filter. The colour maps thus
obtained were examined to identify features like star forming regions, dust
lanes, etc. and study their photometric properties and locations with
respect to the underlying galaxy. See Fig. \ref{img}a-j. Grey scales
have been chosen to maximize the contrast over the range of colour
indices.

The surface brightness distribution and the variation of the position angle
and ellipticity of the isophotes of each galaxy were obtained using the
ISOPHOTE package within STSDAS\footnote{The Space Telescope Science Data
Analysis System STSDAS is distributed by the Space Telescope Science
Institute.}. Ellipses were fit to the galaxy images in all the four pass
bands after masking out any foreground stars. The ellipse fitting was done
using the algorithm proposed by Jedrejewski (\cite{jed}). The deviation of
an isophote from a perfect elliptical shape can be estimated by expanding
the difference in intensity between the isophote and the corresponding
fitted ellipse as a Fourier series in the eccentricity anomaly along the
isophote. A negative coefficient B4 of the cos(4$\phi$) term indicates
boxiness while a positive coefficient indicates the presence of a disk.
Though this technique was initially used in the analysis of elliptical
galaxies, in recent years, it has been successfully used in identifying
structures like bars, isophotal twists, dust lanes, etc. in disk galaxies
also (Wozniak et al. \cite{woz95}, Wozniak \& Pierce \cite{woz91}). Analysis
of a sample of MBG starbursts has been carried out by Barth et al.
(\cite{barth}) using ellipse fitting techniques.
Non-circular motions of the gas increase the rate
of collisions, as well as the radial transport of gas to the galactic
center and are seen as isophotal twists. Scoville \& Hersh (\cite{scov}) suggest that this mechanism
could be responsible for enhancing the gas density in the nucleus and 
triggering star formation there.
Generally, all the parameters of the ellipses, viz ellipticity, position
angle and center, were allowed to vary. However, in Mrk 743, which shows a
double nucleus, the center was kept fixed at a value derived from the
outermost isophotes. Ellipses were fit right up to the central pixel in all
cases except in Mrk 743 due to its double nucleus and in Mrk 363 which does
not show a well defined nucleus in the B band. For these galaxies, the
minimum radius of fitting was decided on the basis of the extent of the
central structure. The results of the ellipse fitting viz. the radial
variation of the surface brightness, the ellipticity (e) and the position angle (P.A.)
are presented in Fig. \ref{sep}.a-i.  The variation in colour is
presented in Fig. \ref{col}. From the ellipticity of the outermost
fitted isophotes, we derive the angle of projection of the galaxy on the
plane of the sky using the equation given by Tully (\cite{tully}). 
\begin{eqnarray}
i &=& 
cos^{-1}
{
\left\{
{{(b/a)^2 - 0.2^2}\over{1-0.2^2}}
\right\}
}^{1/2}
+ 3\degr\;
\nonumber\\
e &=& 1-{b\over a}
\nonumber
\end{eqnarray}

The results are tabulated in Table 1.
The total
light enclosed within the isophote
at 24, 23.5, 23.5 and 22.5 mag/\arcsec$^{2}$ in the B, V, R and I images
respectively, was used to determine  m$_B$, m$_V$, m$_R$ and m$_I$ by summing over the light in the fitted elliptical annuli.
The magnitudes derived in this manner and colour indices for each galaxy are
presented in Table 3. along with the spatial distribution of the starburst. 

In order to identify non axisymmetric structure and other small scale
features, unsharp masked images were constructed for each galaxy. For this,
we produced models for the smooth light distribution of the galaxies in
each band pass using the parameters derived from the isophotal analysis.
This smoothed image was subtracted from the original image, to produce a
residual image with enhanced features not apparent in the direct images.
Such fine structure provides a wealth of information about the processes at
play in the galaxy. We detect complex fine structure most prominently in
the galaxies Mrk 1002, Mrk 363 and Mrk 213. For a detailed discussion refer
to section 4. Schweizer \& Seitzer (\cite{ss88},\cite{ss92}) quantify such
fine structure in E and S0 galaxies that are believed to be induced by
mergers. Simulations by Hernquist (\cite{hern92},\cite{hern93}), also
predict features like boxy isophotes and X-shaped fine structure for merger
products.
\begin{figure*}
\vspace{1cm}
\caption[]{Radial distribution in the colour indices for the galaxies.}
\label{col}
\end{figure*}

\begin{table*} 
\caption[] {Total magnitudes and colour indices uncorrected
for galactic absorption and inclination effects. The last column describes
the locations of the starburst, where nuc : nuclear, comp : companion galaxy}
\begin{tabular}{llllllllllll}\hline
Mrk &\multicolumn{2}{c}{B$_{T}$}&\multicolumn{2}{c}{V$_{T}$}&R$_{T}$&I$_{T}$&m$_{B}$ &m$_B$-m$_V$& m$_V$-m$_R$ & m$_R$-m$_I$&starburst\\
&this work&RC3&this work&RC3&&&&&&&\\
 \hline
14 &14.67&14.9&14.11&14.49&13.54&13.09& 14.74 & 0.52 & 0.51 & 0.47&nuc  \\
87 &13.87&13.8&12.77&12.85&12.04&11.49& 14.12 & 0.84 & 0.67 & 0.71& nuc+ring\\
213 &13.18&13.10&12.68&12.50&12.06&11.44&13.39&0.60 & 0.54& 0.59&nuc+bar end\\
363&14.31&14.28&13.76&13.72&13.20&12.81& 14.34 & 0.52 & 0.56 & 0.32&global \\
449 &13.45&13.50&12.87&12.85&12.39&11.69& 13.69 & 0.66 & 0.53 & 0.63&nuc \\
743 &13.19&-&13.04&-&12.67&12.06& 13.57 & 0.36 & 0.47 & 0.63&one nucleus \\
781 &13.07&13.19&12.42&-&11.92&11.44& 13.40 & 0.64 & 0.53 & 0.59&nuc+spiral arms \\
1002 &13.60&-&13.05&-&12.47&12.09& 13.72 & 0.57 & 0.55 & 0.60&clumpy nuc \\
1308 &14.04&-&13.45&-&12.99&12.39& 14.24 & 0.50& 0.47& 0.52&nuc\\
1379 &13.29&13.00&12.59&12.34&12.04&11.31& 13.43 & 0.63 & 0.58 & 0.77&nuc+bar end+comp \\ \hline
\end{tabular}
\end{table*}		
\section{Notes on individual galaxies}

1. {\bf Mrk 14} : This galaxy has been classified as likely to be an S0
by Huchra (\cite{huchra}), while it forms a part of the sample of distant
irregulars in a study conducted by Hunter \& Gallagher (\cite{hunt}).  
The contours in the outer regions look disturbed. See Figs. \ref{img}a.
and \ref{sep}a. The ellipticity profiles show a complex structure.
Ellipticity is highest ($\approx$0.3) in the central region in all the
filters. It drops to a value of 0.1 at 9\arcsec\ . Beyond 9\arcsec\, the
ellipticity rises again, showing a number of peaks before reaching a
value of 0.2 in the outermost region. The isophotes show a continuous
twist (40\degr) from the central to the outer regions. The contours in
the inner 8\arcsec\ show a boxy nature while beyond 10\arcsec\, they
become pointy as is seen from the negative and the positive values
respectively of the coefficient B4 of the cos(4$\phi$) term depicted in
Fig. \ref{m14b4}. Keel \& van Soest (\cite{keel}) find no candidate
companions near this galaxy. The only significant feature in the colour
maps is the blue nucleus (Fig. \ref{img}a.). The (B-I) colour map does not
show any features like dust lanes which can be responsible for the strong
isophotal twist. The variation of the colour indices, (V-I), (B-R) and
(B-V) with distance from the center are shown in Fig. \ref{col}. On the
whole, the colours get redder outwards. A sharp change is seen in the
inner 8\arcsec\ . (B-V) changes steeply from 0.15 to 0.75 in this region.
Beyond this, the change is more gradual. The residual image constructed with
the model light distribution subtracted from the direct image fails to
reveal any other structure besides a bright nuclear region. A strong isophotal twist and
boxiness of the isophotes are indicative of possible interaction in the
past. Similar inferences have been drawn by Nieto \& Bender
(\cite{nieto}) and Bender \&  M\"{o}llenhoff (\cite{bender}) for other
early-type galaxies. This is the
most likely cause for triggering the central starburst in Mrk 14. Besides interactions in the past, an alternative scenario has been suggested by the referee
which is also capable of explaining the observed features like the
peak in ellipticity, the isophotal twisting and boxiness. He suggests that
the above features could also arise from a triaxial bulge remnant of a bar
destroyed by the gas accumulation in the nucleus. For such a scenario, the colour gradients can be explained by the star formation that occurred during the
gas inflow, the twisting of isophotes could be due to the triaxial bulge and
the boxiness as well as the ellipticity peak could be a signature of the now
disappeared bar. Hence, the observed features in Mrk 14 could be expected
to arise either as a result of a past interaction or the formation of a triaxial bulge from a bar due to mass inflow towards the center.
\begin{figure}
\vspace{1cm}
\caption{The B4 coefficient as a function of the semi-major axis for Mrk 14.}
\label{m14b4}
\end{figure}

2. {\bf Mrk 87} : This is an example of a ringed galaxy. It shows a
bright nucleus and a bar surrounded by an inner ring. The nucleus, the
bar and the inner ring appear prominently in the B band image. Towards
longer wavelengths, the light distribution becomes smoother (Fig. \ref{img}b.).
A companion is situated to the NE of Mrk 87. 
The central regions seem to be disturbed due to the presence of dust as
indicated by the filter dependent behaviour of the ellipticity and the
position angle in the inner 5\arcsec\ in Fig. \ref{sep}b. Beyond 5
\arcsec\ , the behaviour is similar in all filters. The ellipticity
profiles show a slight discontinuity at 8\arcsec\ . Between 5\arcsec\ and
10\arcsec\ , the position angle in nearly constant at 75\degr\ and then
slowly changes by 15\degr\ before reaching the final value of 60\degr\ .
The presence of the blue ring is also clearly seen in Fig. \ref{col} at
24\arcsec\ . Rings of stars and gas are often seen in barred spiral
galaxies. They are believed to be formed by gas accumulation at the bar's
Lindblad resonances. They are blue in colour and are the sites of
enhanced star formation (Buta \cite{buta}). Arsenault (\cite{arsenault})
has shown that the bar and ring features occur with a higher frequency in
starburst galaxies as compared to normal galaxies.

3. {\bf Mrk 213} :  This is a barred spiral galaxy. A faint arm is seen
emerging from the south-eastern end and curving around towards the north-western
side where it is attached to an almost stellar condensation. The
contour plots (Fig. \ref{img}c) clearly show that the contours in the
inner regions have a position angle different from that of the outer
region. The contours in the central region appear elliptical. A very bright, saturated star
is present at 58\arcsec\ from the center of the galaxy. This star was first masked out
before starting the ellipse fitting procedure. The
ellipticity profiles show a peak at 4\arcsec\ . 
Beyond this, the ellipticity falls and then rises again to the disk ellipticity of
about 0.5. The position angle curves show a nearly constant value in the inner
8\arcsec\ . We attribute these features to the presence of a nuclear bar
in the central region. Such nuclear bars have been detected by Jungwiert
et al. (\cite{jung}) and references therein and Wozniak et al.
(\cite{woz95}) in a number of disk galaxies. These bars are believed to
be an efficient mechanism for driving gas into the nuclear region and
fueling the starburst. The position angle changes by 25\degr\ between
8\arcsec\ and 10\arcsec\ 
(Fig. \ref{sep}c.).
A dust lane starts from the nucleus, curves around it
before proceeding towards the NW direction. The residual maps also show
the signature of the bar in the form of a linear structure in the central
region(See Fig. \ref{img}c.).

4. {\bf Mrk 363} : This peculiar Sc type galaxy has been classified by
Geller \& Huchra (\cite{geller}) as belonging to a group made up of seven
galaxies, based on their proximity in space and their radial velocities.
A neutral hydrogen mapping by van Moorsel (\cite{van}) shows a central
concentration of HI. Radio emission extended over the galaxy has been
observed by Wrobel \& Heeschen (\cite{wrobela}). An inspection of the contours in the
B and the I band (Fig. \ref{img}d.) reveals that the morphologies are
quite different in the two spectral windows. 
The blue continuum has an extended structure in the central region with
no well defined nucleus while the I band image shows a well defined
nucleus in the central region. The contours appear highly disturbed and
asymmetric in both the bands, though they are more so in the B band. The
contours in the B band are stretched out in the north more than those in
the I band suggesting the presence of dust in the region at $\approx$
9\arcsec\ from the nuclear region. The (B-I) map clearly shows a highly
reddened region coinciding with this feature. 
This region has a mean (B-V) of 0.7 while the nuclear region has a mean
value of 0.3. The I band image shows a pointed structure starting from
the nuclear region and extending up to 4\arcsec\ along the SE direction.
There is another pointy structure starting at 6\arcsec\ and extending
up to 12\arcsec\ along the southern direction. 
The ellipticity in the central and the outer regions is 0.2 as is seen
from Fig. \ref{sep}d. Close to the nucleus, the ellipticity profile shows
a small kink. It rises to a value of 0.4 at 9.5\arcsec\ - the region
where the contours start becoming pointy again. Subtracting the smoothed
image from the direct images revealed complex fine structure. A bright
and stellar nucleus, with spiral arms lying along the north-south
direction and extending right into it are seen in Fig. \ref{img}d. As this
galaxy belongs to a group of seven galaxies, tidal interactions with the
other galaxies of the group are a likely cause of the enhancement in star
formation activity in this galaxy.

5. {\bf Mrk 449} : This is the most inclined galaxy in our sample. We
derive an inclination of 75\degr\ for this galaxy. The contour maps show
the presence of a highly reddened region lying to the east of the central
nucleus. This feature gets weaker towards longer wavelengths.
Correspondingly, the colour map also shows a reddened vertical band in
this region (refer Fig. \ref{img}e.). We interpret this as a dust lane.
This dust lane lies neither along the major axis of the galaxy, nor along
its minor axis, but is at an intermediate angle. Hawarden et al.
(\cite{haward}) observed such "skew" dust lanes in a small fraction of
early type galaxies. 
They suggest that the properties of the galaxies with such "skew" dust
lanes are best attributed to the accretion of cool material at a fairly
recent epoch. The signature of this is also seen in the surface
brightness profiles as a dip at 10\arcsec\, most prominent in B and
weakest in I (Fig. \ref{sep}e.), and as a reddened region in the colour
plots in Fig. \ref{col}. A cut across the dust lane taken at a position angle
97\degr\ is presented in Fig.
\ref{cbr}, which clearly shows the extent and the reddening due to the
dust lane in (B-R). (B-R) changes from 0.8 in the nuclear region to 1.5
at the location of the dust lane. The position angle remains constant
throughout except for a small dip in the central region. The very high
inclination of this system makes it difficult to interpret the results of
the ellipse fitting uniquely. The residual image shows complex fine
structure. Excess of luminosity is present in the nucleus as well as in
the region to the east of the dust lane.

\begin{figure}
\vspace{1cm}
\caption{Variation in (B-R) along a cut through the nucleus and the dust lane in Mrk 449 at a position angle of 97\degr.}
\label{cbr}
\end{figure}

6. {\bf Mrk 743} : is classified as a peculiar E0 galaxy in the Markarian
catalog. It forms a part of the sample of galaxies with double nuclei
studied by Mazzarella \& Boroson (\cite{mazz93}). An inspection of the
contour plots in the four bands reveals the presence of two nuclei
surrounded by a common envelope. The envelope is asymmetric in the outer
regions. Both the nuclei have comparable fluxes in the V band. As we go
towards longer wavelengths, the western nucleus starts getting brighter,
while the eastern component becomes
dominant at shorter wavelengths (Fig. \ref{img}f.). 
Mrk 743 is one of the few HI sources among early type galaxies in which
the HI distribution shows a central concentration, rather than the usual
depression (Burstein et al. \cite{burstein}). The HI distribution is in
the form of a disk nearly as large as the galaxian diameter (van Driel \&
van Woerden \cite{van91}). Wrobel \& Heeschen (\cite{wrobelb}) detect
unresolved radio emission at 6cm from this object. We fitted ellipses to
this galaxy keeping the center coordinates fixed at a point between the
two nuclei. 
This was maintained for all the filter bands. Ellipses were fit only in
the common envelope region. The surface brightness profile is shown in
Fig. \ref{surf}. The colour plots in Fig. \ref{col} show a sharp change
in the inner 10\arcsec\ . Since Mrk 743 shows such a peculiar morphology,
one cannot comment on the nature of the underlying galaxy from the
results of the ellipse fitting process. However, the presence of two
nuclei with star formation enhanced in only one and the presence of an
asymmetric outer envelope all indicate that this is a merger in progress.

\begin{figure}
\vspace{1cm}
\caption{Surface brightness distribution of Mrk 743 in B (filled
squares),V (filled hexagons),R (crosses) and I (triangles).}
\label{surf}
\end{figure}

7. {\bf Mrk 781} : This is is a barred galaxy with a flocculent spiral
structure. No conspicuous difference is seen in the structure between the
B and I wave bands in Fig. \ref{img}g. We detect a blue ring
surrounding the nuclear region in the colour images. Besides the ring
and the blue spiral arms, the rest of the galaxy has (B-V) between 0.8-0.9. The extent of the
bar is estimated from the peak in the ellipticity profiles in Fig.
\ref{sep}g. to be 30\arcsec\ . However, there is a systematic shift in
the peak ellipticity from B to I with the peak in B having the largest
ellipticity. This is due to the star forming regions at the end of the
bar which contribute at shorter wavelengths.
Resonances created by the bar are possibly
responsible for the star forming ring in the central region.
The diameter of the ring is found to be 3\arcsec\ . Faint signatures of the ring are also
seen in the ellipticity and the colour profiles. However, since the seeing
was $\sim$ 1.5 \arcsec\ , imaging at a higher resolution is required to study the
details of the ring.

8. {\bf Mrk 1002} : This galaxy is classified as S0 by Mazzarella \&
Balzano (\cite{mazz}) and as E1 by the Markarian catalog. The central
contours, as seen in Fig. \ref{img}h., appear elliptical. An inspection of the
direct images reveals that at about
6\arcsec\ , the contours start deviating from ellipses and seem to give a
faint indication of spiral arms. Beyond 12\arcsec\ , they regain their
elliptical nature. The position angle jumps abruptly between 3\arcsec\
and 10\arcsec\ . 
This is accompanied by a sudden sharp increase in ellipticity from less
then 0.1 in the central region to 0.3 at 10\arcsec\ (Fig. \ref{sep}h.).
This jump can be attributed to the spiral arm like features. Between
5\arcsec\ and 10\arcsec\, there is a twisting of isophotes, accompanied
by boxiness of the isophotes in this region, as seen from the negative
values of the B4 coefficient in Fig. \ref{m1002b4}. We detect a S-shaped
blue structure crossing the nucleus in the color maps. The residual maps
also show this structure partly. The excess luminosity in this structure
shows a one-to-one correspondence with the blue areas in the colour map.
The colour maps and the structural features imply star formation in
certain regions of this galaxy. Pogge \& Eskridge (\cite{pogge}) have
reported H$\alpha$ emission in the nuclear as well as clumps of emission
in the circumnuclear region in this galaxy. The rudimentary spiral
pattern, the isophotal twists and the boxiness all suggest that this is a
possible case of a merger wherein a gaseous disk was captured by this galaxy at
some point. A similar spiral feature is found in the early type galaxy of the interacting pair
AM 0327-285 by de
Mello et al. (\cite{demello})  which they explain as arising due to the
capture of a gaseous disk by the early type galaxy. However, because of the complex nature of the structure, a definitive answer can only be given with additional kinematical information on this galaxy.

\begin{figure}
\vspace{1cm}
\caption{The B4 coefficient as a function of the semi-major axis for Mrk 1002.}
\label{m1002b4}
\end{figure}

9. {\bf Mrk 1308} : This is a small nearby galaxy of S0 type extending
about half an arcminute. It has a small linear companion located at
30\arcsec\ towards the west, which has been confirmed to be a physical
neighbour by 6m telescope spectroscopy (Doublier et al. \cite{doub}). The
contours in the B and I filter bands appear smooth (see Fig. \ref{img}j.).
However, we detect a very strong systematic twisting of isophotes in this
galaxy (Fig. \ref{sep}h.). Though the ellipticity shows a small variation
of less than 0.1 beyond the inner 3\arcsec\, there is a continuous
variation in the position angle. Within the inner 10\arcsec\ , the
position angle changes by nearly 180\degr. Mrk 1308 exhibits strong
isophotal twisting overall and non-concentric isophotes in the outer
regions. The B4 coefficient oscillates around zero and shows no clear
trends.
Besides the blue nuclear region, there are no other features detectable
in the colour maps. The (B-V) colour for Mrk 1308 is 0.05 in the central
regions and gets redder outwards, reaching a value of 0.7 near the
periphery. The companion is a red object having a mean (B-V) of 0.9. The
star formation activity is confined to the nuclear region. Radio imaging
at 6cm. (Neff \& Hutchings \cite{neff}) shows that the emission is in
the form of a ring-like structure of diameter $\approx$ 3\arcsec\ . The
residual image obtained after subtracting a smoothed image constructed
using the isophotal analysis, reveals a bright nucleus and another blob
to its north. The tidal interactions induced by the close companion could be
responsible for the non-axisymmetric perturbations seen in the central part of
Mrk 1308. The gas inflow to the nucleus and hence the star formation in
the nucleus could arise as a result of these
perturbations.
seem
to be the source of the starburst phenomenon in this galaxy.

10. {\bf Mrk 1379} : is a VV object (Vorontsov-Velyaminov \cite{vv}), with
nests of interacting objects. The irregular galaxy lying toward the
eastern edge of Mrk 1379 was first masked out to prevent the results of
ellipse fitting to be influenced by its luminosity.  The position angle
increases from 60\degr\ to 75\degr\ between 5\arcsec\ and 8\arcsec\
accompanied by a kink in the ellipticity profile between these points. We
detect two peaks in the ellipticity profiles, the ellipticities of which
are wavelength dependent. The first one corresponds to the blue structure
lying at 13\arcsec\ to the west of the nucleus. Also, the local minima at
19\arcsec\, between the two maxima becomes more prominent at longer
wavelengths. 
There is no appreciable variation in the position angle in these two
regions. However, the surface brightness profiles have a small kink at
this position in all the filters, the most prominent being in the B band. 
In the inner 10\arcsec\ the blue isophotes are rounder. Between 10\arcsec\
and 15\arcsec\ the ellipticity profiles do not show any wavelength dependence.
Beyond 15\arcsec\ the trend reverses and the I band profiles are rounder. Beyond
30\arcsec\ the contribution to the blue light is predominantly from the spiral
arms while the light distribution at longer wavelengths is smoother and in the form of a common envelope, which accounts for the wavelength dependent behaviour of the ellipticity
profiles in this region.
The (B-I) image shows
nuclear as well as extra nuclear star formation. Knots of star formation
are seen in the nuclear region as well as along the western periphery of
the galaxy at the point where the spiral arms start. In addition to this,
global star formation is detected in both the companions lying to the
east. 
The residual maps in the B and the I band appear different. Enhancements
are seen in the nuclear region, in the blue knot to the west and along a
curved spiral arm which appears to start from the nucleus. However, on
inspection of the fine structure in the I band, we clearly detect a short
bar in the central region. We also detect spiral arms with two bright
blobs connected to this inner linear structure.
\begin{figure*}
\vspace{1cm}
\caption[]{Exponential fit to the R bandpass luminosity profiles}
\label{disk}
\end{figure*}
\begin{figure}
\vspace{1cm}
\caption{Comparison of the scale lengths in B ($h_{b}$) and R in ($h_{r}$). The
points are the derived values and the solid line is the locus of
$h_{b}$=$h_{r}$.}
\label{scale}
\end{figure}
\begin{figure}
\vspace{1cm}
\caption{Comparison of the half-light radius in B ($a_{eb}$) and in R
($a_{er}$). The points denote the derived values and the solid line is
the locus of $a_{eb}$=$a_{er}$.}
\label{hrad}
\end{figure}
\begin{figure}
\vspace{1cm}
\caption{Plots of the blue central disk surface brightness versus the blue scale length. For comparison with the values obtained by de Jong, the scale lengths have been transformed appropriately.}
\label{err}
\end{figure}
\section{Decomposition parameters} 

The decomposition of a galaxy profile into a bulge and a disk component
requires the profile to be fitted by the sum of two empirical laws. The
presence of structures like bars, dust lanes, lenses and rings render
such a straightforward two component fit very difficult. The luminosity
profiles of the program galaxies are complex in nature. The burst of star
formation in the nuclear region manifests itself as a sharp rise in
intensity in this region. The burst luminosity completely dominates the
light output in this region particularly and the estimation of a bulge
component becomes very difficult. Hence we characterize the light
distribution by parameters like the half-light radius, the central disk
surface brightness and the disk scale lengths.

\subsection{Scale lengths and central disk surface brightness}
	
In the Markarian starburst galaxies, the luminosity profiles are 
complex in nature. In the inner
region, the profile falls steeply up to about 10 \arcsec\ where the light is
completely dominated by the burst component. The outer parts of the
luminosity profile in most cases can be well described by an exponential
scaling law viz.
\begin{eqnarray}
\mu(r)& =& \mu_0 + 1.086{\left(r\over h\right)}.
\nonumber
\end{eqnarray}
The outer exponential nature is also seen in case of dwarf ellipticals
dwarf irregulars and HII galaxies (Telles \cite{telles}). This outer
part is likely to represent the old underlying population of the parent
galaxy. Structural properties like scale lengths and central disk surface
brightnesses can be derived from these profiles. We fit a exponential law
to the outer part of the profile down to where the signal falls to
3$\sigma$ of the background noise level. Fig. \ref{disk} depicts
the fits to the outer regions. Good fits to the observed profiles are obtained
for the galaxies Mrk 14, Mrk 743, Mrk 1002, Mrk 1308 and Mrk 1379.
In case of Mrk 449, the luminosity profile shows a sharp dip due to the
presence of the dust lane and hence the fitted values are higher than the observed values
for surface brightness in this region. We have deliberately included a part of the bar into the range of fit for Mrk 213, Mrk 781 and Mrk 1379 since we found that excluding the bar region tends to overestimate the intensity in the inner regions of the galaxies.
The results of exponential fits to each
filter are tabulated in Table 4. A plot of the scale length in B versus
the scale length in R (Fig. \ref{scale}.) indicates that the blue scale
lengths are comparable to the red scale lengths in all the cases except for Mrk 87.
The exponential law fails to fit very well in the outer regions for Mrk 87 and the departures seen could be a result of this
improper fit. The total magnitudes B$_T$, V$_T$, R$_T$ and I$_T$ were derived by extrapolating
 the fitted disk to infinity and summing over the flux.

\begin{table*}
\caption[]{Distribution of half-light radii ($a_{e}$), scale lengths (h) and the
central disk surface brightnesses ($\mu_{o}$). The subscripts b,v,r,i denote the filters for which the values are presented. The half-light radii and scale
lengths are in kpc ,the surface brightness in mag/sq.arcsec. and the range of fit in arcsec. The first row gives the derived values and the second row gives the errors on the corresponding quantities.}
\begin{flushleft}
\begin{tabular}{llllllllllllll}
\noalign{\smallskip}
\hline
\noalign{\smallskip}
Galaxy&$a_{eb}$&$a_{ev}$&$a_{er}$&$a_{ei}$&$h_{b}$&$h_{v}$&$h_{r}$&$h_{i}$&$\mu_{ob}$&$\mu_{ov}$&$\mu_{or}$&$\mu_{oi}$&range of fit\\
\hline
14&0.99&1.34&1.47&1.47&1.13&1.37&1.29&1.27&20.76&20.39&19.68&19.10&7-21\\
&&&&&0.04&0.02&0.03&0.04&0.08&0.03&0.04&0.07&\\
87&2.8&2.4&2.1&2.5&7.66&5.89&4.87&4.10&21.92&21.04&20.22&19.39&14-31\\
&&&&&1.23&1.22&0.42&0.22&0.07&0.14&0.07&0.05&\\
213&3.38 &2.75 &2.28 &3.03 &2.77&2.75 &2.60 &2.74 &20.09 &19.61 &19.01 &18.54&13-42 \\
&&&&&0.46&0.35&0.42&0.32&0.19&0.15&0.19&0.15&\\
363&1.55&1.59&1.63&1.84&0.76&0.82&0.78&1.02&18.72&18.55&17.82&18.27&14-21\\
&&&&&0.05&0.04&0.04&0.03&0.26&0.10&0.23&0.19\\
449&1.33&1.33&1.21&1.46&0.71&0.73&0.69&0.66&19.18&18.49&17.89&17.19&14-40\\
&&&&&0.032&0.025&0.019&0.023&0.08&0.06&0.05&0.07\\
743&0.35&0.49&0.40&0.54&0.74&0.45&0.48&0.57&21.10&20.02&19.60&19.25&7-21\\
&&&&&0.024&0.007&0.006&0.011&0.07&0.06&0.04&0.03\\
781&5.30&6.00&5.83&5.85&3.14&3.07&2.96&2.47&21.44&20.71&20.11&19.32&10-40\\
&&&&&0.54&0.21&0.18&0.14&0.04&0.03&0.03&0.03\\
1002&2.28&2.28&2.51&2.51&1.79&1.78&1.85&1.79&20.06&19.39&18.86&18.38&7-22\\
&&&&&0.045&0.137&0.098&0.090&0.16&0.11&0.10&0.11\\
1308&0.38&0.53&0.56&0.65&0.36&0.39&0.40&0.43&20.02&19.44&18.95&18.45&7-21\\
&&&&&0.008&0.005&0.012&0.011&0.04&0.03&0.04&0.03\\
1379&2.32&2.63&2.39&2.83&1.94&1.81&1.72&1.66&20.17 &19.37 &18.72 &17.96&11-35\\
&&&&&0.074&0.054&0.056&0.047&0.05&0.04&0.05&0.04\\
\hline
\end{tabular}
\end{flushleft}
\end{table*}

\subsection{Half-light radius}

The growth curve in each filter band was used to determine the
half-light radius, $a_{e}$ namely the radius within which half of the total
light of the galaxy is contained.

\begin{eqnarray}
m_{\rm hl}& =& m_{\rm T} + 0.7525
\nonumber
\end{eqnarray}
The values for the total magnitudes were
taken from Table 3. to compute the half-light radii.
The half-light radii derived for the sample galaxies in each of the filter
bands are presented in Table 4. Fig. \ref{hrad} shows the plot of
the half light radius in B versus the half-light radius in R. The plot does not show any clear
trends, however there are indications of $a_{eb}$ being smaller than $a_{er}$
in most cases. This suggests that the blue light is more centrally concentrated
then the red light in most of these objects within the surface brightness limits reached by
our data.
This is to be expected in case of starburst galaxies as the starburst
activity is nuclear or circumnuclear in most of the galaxies.

\section{Comparison with other samples}
The derived parameters viz. the scale lengths and the central disk surface
brightnesses were computed with those derived by de Jong (\cite{dej})
for normal spiral
galaxies. The plot of the blue scale length versus the corrected
central disk brightness
(Fig. \ref{err}) shows that the Markarian starburst galaxies in our
sample have scale lengths less than about 2 kpc in the B band. These are
shorter compared to those derived for normal spirals
by de Jong using a bulge-disk decomposition. The central disk surface
brightness is also brighter than that observed in normal spirals by other workers.
Bothun et al. (\cite{bothun}) report a mean value of 2 kpc for
the scale lengths derived using "marking the disk" based on the study
of the Wasilewski sample of emission-line galaxies in the Gunn R band.
They also find the mean central surface
brightness of exponential disks to be about 1.5 mag brighter than that for normal spirals in Gunn R. However, Courteau (\cite{cort}) cautions against the
comparison of scale lengths derived by different workers due to the subjective
nature of the measurements.  He argues that "marking the disk" approach would
lead to smaller scale lengths over the bulge-disk decomposition fits. In the
(B/D) decompositions, the inclusion of a bulge reduces the amount of light
contributed by the inner disk and thus leads to a shallower slope for the
disk profiles. Knapen \& van der Kruit (\cite{kn}) find discrepancies up to a factor of
two in the scale lengths measured by various workers.

\section{Properties of the burst region}

Huchra (\cite{huch}) was the first to propose that the starburst galaxy
can be thought of as made up of stellar populations of two ages : the
underlying galaxy which is the old galaxy, superimposed on which is the
young burst component. To estimate the burst component, the contribution
of the underlying galaxy has to be subtracted. This is not an easy task.
Aperture photometry of the region in which the young population is
present gives the total intensity in that region. The contribution of the
galaxy can be subtracted by estimating the galaxy from an annular region
around the burst region. However, this approach suffers from a major
drawback. The galaxy contribution estimated by this method gives the
underlying galaxy value outside the region and not at the position of the
burst. Such an approach would underestimate the galaxy contribution
especially when the underlying galaxy luminosity profile is an
exponential disk as described in the previous section. This would lead to
erroneous values for the colour of the burst. To overcome this
difficulty, we use the disk component estimated in the last section to
derive the burst values. The exponential disk is extrapolated right up to
the central regions. We construct a disk model of the galaxy. This is
then subtracted from the galaxy. The residual intensity in the central
region gives us the contribution of the burst component. This approach is
used only for the central region in each galaxy since non-axisymmetric
structures like bars can contribute to the residual in the outer regions.
The burst colours derived after subtracting the disk component as described 
above are presented in Table 5. Mrk 363 shows globally enhanced star formation. This makes it impossible to separate the young and the old stellar components
in this galaxy and hence estimate the burst colours from the results of the disk fitting.
\begin{table}
     \caption[]{Colours of the burst component}
	\begin{tabular}{llll}
	\hline
	Galaxy & B-V & V-R & R-I\\
                \hline
	Mrk 14 & 0.33&0.36&0.32\\
	Mrk 87&0.54&0.63&0.55\\
	Mrk 213&0.63&0.63&0.42\\
	Mrk 363&- &- &- \\
	Mrk 449&0.09&0.16&0.111\\
	Mrk 743& 0.38&0.32&0.22\\
	Mrk 781&0.67&0.48&0.38\\
	Mrk 1002&0.37&0.50&0.04\\
	Mrk 1308&-0.04&0.29&0.14\\
	Mrk 1379&0.36&0.47&-\\

	\hline
                \end{tabular}
        \end{table}

\section{Conclusions}

The starburst activity is seen to be hosted by galaxies of nearly all
morphological types. In case of elliptical or S0 galaxies in the sample,
strong isophotal twisting has been detected, accompanied by
boxiness in certain cases. Mrk 1002, Mrk 14 and Mrk 1308 all show strong
isophotal twists. In addition, boxiness is detected in Mrk 1002 and Mrk
14. Fine structure like rudimentary spiral structure is observed in Mrk
1002. Nuclear bars have been detected in Mrk 213 and Mrk 1379. Tidal interactions, mergers or accretion of material are the most
likely triggers for the starburst phenomena observed in S0s and
elliptical like galaxies in the sample. The spirals in the sample show the presence of
a bar responsible for fueling the starburst or the presence of a
bar-like structure in the central nuclear region as in the case of Mrk
213.

\begin{acknowledgements} 
This work was supported by the Department of Space, Government of India.
The authors are thankful to Mr. A.B. Shah, Mr. N.M. Vadher and Mr.
Shashikiran Ganesh for technical support. We are grateful to the referee,
Dr.Wozniak for valuable suggestions. 
\end{acknowledgements}

\end{document}